\documentclass[fp]{jpsj3}
\usepackage{amscd}
\newcommand{\del}{\delta}
\newcommand{\ep}{\epsilon}
\usepackage{txfonts}

\title{Derivation of Markovian Master Equation by Renormalization
  Group Method}
\author{Yasusada nambu$^1$\thanks{E-mail: nambu@gravity.phys.nagoya-u.ac.jp}
  and Shingo Kukita$^2$\thanks{E-mail: kukita@th.phys.nagoya-u.ac.jp}}
\inst{Department of Physics, Graduate School of Science, Nagoya 
University, Chikusa, Nagoya 464-8602, Japan}
\date{October 4, 2016} 
\abst{ We present a derivation of the Markovian master equation by the
  renormalization group method. Starting from a naive perturbative
  solution of the von Neumann equation, the reduced density matrix
  with the coarse grained time steps is obtained using the assumption
  of short correlation time of the bath field. Then by applying the
  renormalization group method, we show that the dependence of the
  specific initial time on the perturbative solution can be removed
  and the Markovian semigroup master equation in the
  Gorini--Kossakowski--Lindblad--Sudarshan (GKLS) form is obtained in
  the weak coupling limit.  }

\begin{document}

\maketitle

\section{Introduction}

The quantum dynamics of an open system cannot in general be
represented in terms of unitary time evolution. One useful method of
investigating the open quantum system is to formulate its dynamics by
means of an appropriate equation of motion for its density matrix;
this is a quantum master equation.  As  one of the typical examples
of the open quantum system, the particle detector
model~\cite{Unruh1976,Birrell1984} is widely accepted as a tool for
exploring the quantum nature of the considering system. It is composed of
detector variables (system) with internal states that interact with
external quantum fields (bath).  This model has been applied to
phenomena of particle creations in curved spacetimes (Unruh effect and 
Hawking radiation). This model is also used to detect the
entanglement of quantum fields by investigating the correlation between
two independent detectors interacting with the quantum
fields~\cite{Reznik2003,Reznik2005,Benatti2010,Cliche2010,Steeg2009,Nambu2013}.

   The state of the detector is obtained by tracing out the field
  degrees of freedoms which interacts with the detector (reduced
  density matrix). As a result, the evolution of the system is
  determined by the dynamical map $\mathcal{E}$, which is not unitary
  in general.
$$
\begin{CD}
 \rho_T(0)=\rho(0)\otimes\rho_B   @>\text{unitary evolution}>>
 \rho_T(t)=U(t,0)[\rho(0)\otimes\rho_B]U^{\dag}(t,0) \\
 @V\mathrm{tr}_BVV      @V\mathrm{tr}_BVV \\
 \rho(0)   @>\text{dynamical map}>> \rho(t)=\mathcal{E}(t)\rho(0)
\end{CD}
$$
To describe physically allowable processes, the dynamical map should
preserve the complete positivity and normalization of probabilities in
the course of evolution.

The dynamical map can be represented by the master equation that
determines the evolution of the system.  In general, the master
equation becomes an integro-differential equation, and the state at the
specific time depends on the past evolution (non-Markovian
nature). However, it is possible to recover the Markovian property of the
master equation by assuming suitable time scales: the decay time of
the bath correlation function $t_B$ is sufficiently shorter than the
relaxation time scale of the system $t_S$.  By combining this assumption
of the time scales with the so-called  secular approximation (rotation
wave approximation), which neglects transitions via system energy
nonconserving processes, it can be shown that the resulting master
equation has the Gorini--Kossakowski--Lindblad--Sudarshan (GKLS) form
and preserves the trace and complete positivity of the state in the course of
evolution~\cite{GKS1976, Lind1976}. It is known that the
Markovian master equation, which preserves the trace and complete
positivity, should have the GKLS form and generates the dynamical
semigroup~\cite{Breuer2002}.

As the derivation of a master equation that does not rely on the
secular approximation, the time coarse graining method was
proposed~\cite{Lidar2001,Schaller2008,Benatti2010,Majenz2013}. By
introducing a coarse graining time scale, which is assumed to be longer
than the decay time scale of the bath correlation function, it is
possible to derive  the GKLS master equation without
assuming the secular approximation.  The purpose of this paper is to
present an alternative derivation of the GKLS master equation
 using the renormalization group method~\cite{Chen1996,Kunihiro1996}.

The renormalization group method is applied as a tool of asymptotic
analysis of differential equations~~\cite{Chen1996,Kunihiro1996}.  The
naive perturbative solution of differential equations often yields
secular terms due to resonance phenomena. The secular terms prevent
us from obtaining approximate but global solutions. The renormalization
group method is one of the techniques to circumvent the
problem. Starting from a naive perturbative expansion, the secular
divergence is absorbed into constants of integration contained in the
zeroth-order solution by the renormalization procedure. The
renormalized constants obey the renormalization group equation.  We
shortly review this method using the following example.  Let us
consider the van der Pol equation
\begin{equation}
  \frac{d^2x}{dt^2}+x=\ep(1-x^2)\frac{dx}{dt},
  \label{eq:vander}
\end{equation}
where $\ep$ is a constant. For a small $\ep$, the naive perturbative
solution up to the order $\ep$ is
\begin{equation}
  x(t)=Ae^{it}+\ep\frac{t}{2}\,A(1-|A|^2)e^{it}+\text{c.c}+\cdots,
\end{equation}
where $A$ is a constant and nonresonant terms such as those proportional to
$e^{\pm 3it}$ are not explicitly written. The perturbation fails beyond
a time scale $\sim 1/\ep$ owing to the secular term, which grows as
$\propto t$. To improve this naive perturbative solution, we introduce
an arbitrary renormalization point $\tau$ by splitting $t$ to
$t-\tau+\tau$ and the renormalized constant $A(\tau)$ by
$A=A(\tau)+\del A(\tau), \del A(0)=0$, where the counter term $\del A$
is chosen to absorb $\tau$. Thus,
\begin{equation}
\del
A(\tau)=A(0)-A(\tau)\equiv-\ep\frac{\tau}{2} A(\tau)(1-|A(\tau)|^2),
\end{equation}
and the naive perturbative solution can be written as
\begin{equation}
    x(t)=A(\tau)e^{it}+\ep\frac{t-\tau}{2}\,A(\tau)(1-|A(\tau)|^2)e^{it}+\text{c.c}+\cdots.
\end{equation}
As the renormalization point $\tau$ is arbitrary, by equating $\tau$
and $t$, we obtain the renormalized solution $x(t)=A(t)e^{it}$ with
$A(t)$ being the solution of the following renormalization group
equation:
\begin{equation}
    \frac{dA}{dt}=\frac{\ep}{2}A(1-|A|^2).
\end{equation}
This is an amplitude modulation equation that represents the slow
dynamics of the system. The renormalized solution gives an
approximated but global solution to Eq.~\eqref{eq:vander} up to $O(\ep)$.

In this paper, starting from a naive perturbative solution of the von Neumann
equation, we rederive the master equation in the GKLS form with the
time coarse graining by the renormalization group method.  We
will show that the time coarse graining by the renormalization
procedure naturally defines the semigroup structure for the
renormalized constant in discretized time steps of the evolution of
the reduced density matrix.  We use the unit in which $\hbar=1$
throughout the paper.

\section{Derivation of Markovian master equation} 

\subsection{Naive solution of the von Neumann equation}
Concerning the derivation of a naive perturbative solution of the von
Neumann equation, we basically follow the presentation by
Benatti {\it et al.}\cite{Benatti2010}.  The total system is
composed of detector variables (system) interacting with quantum
scalar fields (bath).  The total Hamiltonian is
\begin{equation}
 H=H_0^S+H_0^B+\lambda V,
\end{equation}
where $H_0^S$ is the system Hamiltonian, $H_0^B$ is the bath
Hamiltonian and $V$ is the interaction Hamiltonian with
$V=\sum_A\sigma_A\Phi_A$,  where $\sigma_A$ is the system operator and
$\Phi_A$ is the bath field.  The strength of the interaction between the
system and the bath is determined by the coupling constant $\lambda$, 
which is assumed to be small (weak coupling). The total density operator
$\rho_T(t)$ in the Schr\"{o}dinger picture obeys the von Neumann
equation
\begin{equation}
 \frac{d}{dt}\rho_T=-i[H_0+\lambda V,\rho_T],\quad H_0=H_0^S+H_0^B.
\end{equation}
We aim to obtain an equation for the reduced density operator for the
detector system
\begin{equation}
 \rho(t)\equiv\mathrm{Tr}_B\{\rho_T(t)\}.
\end{equation}
First, we introduce the interaction picture of the density operator
defined by
$$
 \tilde\rho_T(t)=U_0^\dag(t,t_0)\rho_T(t)U_0(t,t_0),\quad U_0(t,t_0)
 =e^{-iH_0\,(t-t_0)},
$$
where $t_0$ is an arbitrary initial time.  Then, $\tilde\rho_T$ obeys
\begin{equation}
 \frac{d}{dt}{\tilde{\rho}}_T=-i[\lambda\tilde V(t),\tilde\rho_T],\quad
 \tilde V(t)=U_0^{\dag}(t,t_0)V U_0(t,t_0),
\end{equation}
where $\tilde V(t)$ is the interaction representation of $V$.  We
integrate this equation from $t_0$ to $t$. After two iterations, the
formal solution is given by
\begin{align}
\tilde\rho_T(t)&=
  \tilde\rho_T(t_0)-i\lambda\int_{t_0}^t dt_1[\tilde
V(t_1),\tilde\rho_T(t_0)] \notag \\
  &\qquad\qquad
  -\lambda^2\int_{t_0}^t dt_1\int_{t_0}^{t_1}dt_2
[\tilde V(t_1),[\tilde V(t_2),\tilde\rho_T(t_0)]]+O(\lambda^3). 
\label{eq:naiveSol}
\end{align}
We assume that during the evolution, the state of the total
system is factorized
\begin{equation}
 \tilde{\rho}_T(t)\approx{\tilde\rho}(t)\otimes\rho_B,
\end{equation}
where $\rho_B=\rho_B(t_0)$. This assumption is justified because the
interaction between the system and the bath is weak and the
correlation between the system and the bath can be neglected when the
bath time scale $t_B$ is shorter than the system time scale $t_S$. As
the bath contains an infinite number of degrees of freedom, the back
action of the system on the bath is negligible and the evolved total
state can be expressed as the product state (11).  By taking the trace
of the perturbative solution \eqref{eq:naiveSol} with respect to the
bath degrees of freedom, the reduced density operator for the system
is
\begin{align}
  \label{eq:naiveSol2}
  \tilde\rho(t)&
  =\rho(t_0)-\lambda^2
\int_{t_0}^tdt_1\int_{t_0}^{t_1} dt_2\mathrm{Tr}_B\left\{
  [\tilde V(t_1),[\tilde
  V(t_2),\rho(t_0)\otimes\rho_B(t_0)]]\right\},
\end{align}
where we have assumed
\begin{equation}
 \label{eq:vev}
 \mathrm{Tr}_B\{\tilde V(t)\rho(t_0)\otimes\rho_B(t_0)\}=0.
\end{equation}
If we do not consider the back action of the system on the bath and
the bath variable evolves freely, the bath operators in the
interaction picture are the same as those in the Heisenberg picture
$\tilde\Phi_A(t)=\Phi_A(t)$. Thus, the condition \eqref{eq:vev} is
equivalent to $\langle\Phi_A\rangle=0$ because
$$
  \mathrm{Tr}_B\{\tilde
  V(t)\rho(t_0)\otimes\rho_B(t_0)\}
  =\sum_A\tilde\sigma_A(t)\langle\Phi_A(t)\rangle=0,
$$
where the expectation value of the bath field $\Phi(t)$ (Heisenberg operator)
is evaluated as
$$
\langle
\Phi(t)\rangle\equiv\mathrm{Tr}_B\{\Phi(t)\rho_B(t_0)\}.
$$

We rewrite the integral in Eq.~\eqref{eq:naiveSol2}. Let us define
\begin{equation}
  A_{12}\equiv\mathrm{Tr}_B\left\{\left[\tilde V(t_1),\left[\tilde
        V(t_2),\rho(t_0)\otimes\rho_B(t_0)\right]\right]\right\},
\end{equation}
and  decompose the integral as
\begin{equation}
  \int_{t_0}^{t}dt_1\int_{t_0}^{t_1}dt_2A_{12}=
  \frac{1}{4}\int_{t_0}^{t}dt_1\int_{t_0}^{t}dt_2(A_{12}+A_{21})
  +\frac{1}{2}\int_{t_0}^{t}dt_1\int_{t_0}^{t_1}dt_2(A_{12}-A_{21}).
\end{equation}
From now on, we omit the symbol $\otimes$ for simplicity. The first integral is
\begin{equation}
  \int_{t_0}^{t}dt_1\int_{t_0}^{t}dt_2\mathrm{Tr}_B\left(\frac{1}{2}\{\tilde
    V(t_1)\tilde V(t_2),\rho(t_0)\rho_B(t_0)\}-\tilde
    V(t_1)\rho(t_0)\rho_B(t_0)\tilde V(t_2)\right),
\end{equation}
and the second integral is
\begin{align}
  &\frac{1}{2}\int_{t_0}^{t}dt_1\int_{t_0}^{t_1}dt_2\mathrm{Tr}_B\left[ [\tilde
    V(t_1),\tilde V(t_2)],\rho(t_0)\rho_B(t_0)\right] \notag \\
  &=\frac{1}{2}\int_{t_0}^{t}dt_1\int_{t_0}^{t}dt_2\,\mathrm{sgn}(t_1-t_2)
    \left[\mathrm{Tr}_B(\rho_B(t_0)\tilde V(t_1)\tilde V(t_2)),\rho(t_0)\right].
\end{align}

Using $V=\sum_A\sigma_A\Phi_A$ and introducing the correlation function of the
bath field by\cite{foot}
\begin{equation}
 G_{A_1\!A_2}(t_1-t_2)=\mathrm{Tr}_B\{\Phi_{A_1}(t_1)\Phi_{A_2}(t_2)\rho_B(t_0)
 \}=\langle\Phi_{A_1}(t_1)\Phi_{A_2}(t_2)\rangle,
\end{equation}
we can rewrite the integral in the right-hand side of
Eq.~\eqref{eq:naiveSol2} as
\begin{align}
  \tilde\rho(t)-\tilde\rho(t_0)
&=
  \Delta
  t\times\bigl(-i[H_{12},\rho_0]+\mathcal{L}[\rho_0]\bigr),
  \qquad \Delta t=t-t_0,\quad \rho_0=\rho(t_0), \label{eq:naive}
\end{align}
where we have defined
\begin{align}
  &H_{12}=-i\frac{\lambda^2}{2\Delta t}\sum_{A_1,A_2}
  \int_{0}^{\Delta t}ds_1\int_{0}^{\Delta t}ds_2\,
  \mathrm{sgn}(s_1-s_2)G_{A_1\!A_2}(s_1-s_2)\,\sigma_{A_1}\!(s_1+t_0)
  \sigma_{A_2}\!(s_2+t_0), \label{eq:H}
    \\
   &\mathcal{L}[\rho_0]=\frac{\lambda^2}{\Delta t}
   \sum_{A_1,A_2}\int_{0}^{\Delta t} ds_1\int_{0}^{\Delta t}
  ds_2 G_{A_1\!A_2}(s_1-s_2)\biggl(
    \sigma_{A_2}(s_2+t_0)\rho_0\sigma_{A_1}\!(s_1+t_0) \notag\\
    &\qquad\qquad\qquad\qquad\qquad\qquad\qquad\qquad\qquad\qquad
    -\frac{1}{2}\left\{
      \sigma_{A_1}\!(s_1+t_0)\sigma_{A_2}(s_2+t_0),\rho_0\right\}\biggr).
  \label{eq:L}
\end{align}
In Eqs.~\eqref{eq:H} and \eqref{eq:L}, the system variable
$\tilde\sigma_{A}(t)$ in the interaction picture is replaced by
$\sigma_{A}(t)$ in the Heisenberg picture whose evolutions are
determined by the free system Hamiltonian $H_0^S$. This is justified
within perturbation up to $O(\lambda^2)$.  The time dependence of the
system variable is determined  by the free Hamiltonian $H_0^S$ as
\begin{equation}
    \sigma_A(t)=e^{iH_0^S(t-t_0)}\sigma_A\, e^{-iH_0^S(t-t_0)}=
    \sum_{B}u_{A\!B}(t-t_0)\,\sigma_{B},\quad u_{A\!B}(0)=1,
\end{equation}
where the specific form of the function $u_{AB}(t)$ depends on the system
Hamiltonian. Using this relation, we obtain
\begin{align}
  &H_{12}=\sum_{B_1,B_2}H_{B_1\!B_2}\sigma_{B_1}\sigma_{B_2}, \\
  &\mathcal{L}[\rho_0]=\sum_{B_1,B_2}C_{B_1\!B_2}
  \left(\sigma_{B_2}\rho_0\sigma_{B_1}-\frac{1}{2}\left\{
      \sigma_{B_1}\sigma_{B_2},\rho_0\right\}\right), 
\end{align}
with
\begin{align}
  &H_{B_1\!B_2}=-i\frac{\lambda^2}{2\Delta t}\sum_{A_1,A_2}
  \int_{0}^{\Delta t}ds_1\int_{0}^{\Delta t}ds_2\,
  \mathrm{sgn}(s_1-s_2)G_{A_1\!A_2}(s_1-s_2)\,u_{A_1\!B_1}(s_1)u_{A_2\!B_2}(s_2),
    \label{eq:HBB}
   \\
   &C_{B_1\!B_2}=\frac{\lambda^2}{\Delta t}
   \sum_{A_1,A_2}\int_{0}^{\Delta t} ds_1\int_{0}^{\Delta t}
  ds_2 G_{A_1\!A_2}(s_1-s_2)u_{A_1\!B_1}(s_1)u_{A_2\!B_2}(s_2). \label{eq:CBB}
\end{align}
As Eq.~\eqref{eq:naive} determines $\tilde\rho(t)$ from the specific
initial state at $t_0$, we cannot obtain the master equation even
though if we take the limit $\Delta t\rightarrow 0$. To convert
\eqref{eq:naive} to the Markovian master equation, we introduce the
time coarse graining and assume the short correlation time of the
bath. Then we apply the renormalization group method~\cite{Chen1996,
  Kunihiro1996} to recover the Markovian master equation.

\subsection{Time coarse graining and renormalization group}
The naive perturbative solution \eqref{eq:naive} has the following structure:
\begin{align}
  &\tilde\rho(t)-\tilde\rho(t_0) \notag \\
&
=\lambda^2\Delta t\times
  \frac{1}{\Delta t}\sum_{A_1,A_2,B_1,B_2}\int_{0}^{\Delta t} ds_1
  \int_{0}^{\Delta t}ds_2\,G_{A_1\!A_2}(s_1-s_2)u_{A_1\!B_1}(s_1)
  u_{A_2\!B_2}(s_2)F_{B_1\!B_2}[s_1-s_2,\rho_0] \notag \\
&
\equiv \lambda^2\Delta t\times\mathcal{D}[\rho_0], \label{eq:naive2}
\end{align}
where $F_{B_1\!B_2}$ represents a combination of the following terms:
$$
  \mathrm{sgn}(s_1-s_2)\sigma_{B_1}\sigma_{B_2}\rho_0,~ 
 \mathrm{sgn}(s_1-s_2)\rho_0\sigma_{B_1}\sigma_{B_2},~ 
 \sigma_{B_1}\sigma_{B_2}\rho_0,~
 \rho_0\sigma_{B_1}\sigma_{B_2},~ 
 \sigma_{B_2}\rho_0\sigma_{B_1}.
$$
Let us introduce a coarse graining time scale $\Delta$ and consider
the evolution of the state by this time step.  Using
\eqref{eq:naive2}, the evolution of $n$
time steps from $t_0$ to $t_0+n\Delta$ is
\begin{align}
  &\tilde\rho(t_0+n\Delta)-\tilde\rho(t_0) \notag\\
  &
  =\lambda^2\Delta\times
    \frac{1}{\Delta}\int_{0}^{n\Delta}ds_1\int_0^{n\Delta}ds_2\,
    \sum_{A_1,A_2,B_1,B_2}\!\!\!\!\!\!G_{A_1\!A_2}(s_1-s_2)\,u_{A_1\!B_1}(s_1)
    \,u_{A_2\!B_2}(s_2)F_{B_1\!B_2}[s_1-s_2,\rho_0]\notag\\
  &
  \equiv\lambda^2\Delta\times\mathcal{D}_n[\rho_0]. \label{eq:evo}
\end{align}
At this point, we assume that the coarse graining step size $\Delta$
is sufficiently larger than the correlation time $t_B$ of the bath
field and the relation $t_B\ll\Delta$ holds. By introducing new
integration variables as $\tau_1=s_1-s_2, \tau_2=s_2-s_1$, the
integral of \eqref{eq:evo} is
\begin{align}
  &\int_0^{n\Delta}d\tau_1\int_{\tau_1}^{n\Delta}ds_2\sum
    G(\tau_1)u(\tau_1+s_2)u(s_2)F[\tau_1] \notag \\
  &\qquad\qquad\qquad\qquad\qquad
    +\int_0^{n\Delta}d\tau_2\int_{\tau_2}^{n\Delta}ds_1
\sum G(-\tau_2)u(s_1)u(\tau_2+s_1)F[-\tau_2] \notag \\
  &\approx\int_0^\infty
    d\tau\int_0^{n\Delta}ds\sum\left(G(\tau)u(s)u(s)F[\tau]
    +G(-\tau)u(s)u(s)F[-\tau]\right),
\end{align}
where we have extended the upper bound of the $\tau_1,\tau_2$
integrals to infinity because the correlation function decays rapidly
for $\tau_1,\tau_2\neq 0$ and the main contribution to the integrals
comes from the region $\tau_1,\tau_2\approx 0$.  Then, $\mathcal{D}_n$
is reduced to the following form:
\begin{equation}
 \mathcal{D}_n[\rho_0]=\frac{1}{\Delta}\int_0^{n\Delta}ds
  \sum_{A_1,A_2,B_1,B_2}\,u_{A_1\!B_1}(s)\,u_{A_2\!B_2}(s)
  \int_{-\infty}^{+\infty}d\tau G_{A_1A_2}(\tau)F_{B_1\!B_2}[\tau,\rho_0],
\end{equation}
and it is possible to  derive the following relations for $n\ge1$:
\begin{align}
(\mathcal{D}_n-\mathcal{D}_{n-1})[\rho_0]&
=\frac{1}{\Delta}\int_{(n-1)\Delta}^{n\Delta}ds
  \sum u(s)\,u(s)\int_{-\infty}^{+\infty}d\tau
 \,G(\tau) F[\tau,\rho_0] \notag \\
 &=\frac{1}{\Delta}\int_{0}^{\Delta}ds
  \sum u(s+(n-1)\Delta)\,u(s+(n-1)\Delta)\int_{-\infty}^{+\infty}d\tau
 \,G(\tau) F[\tau,\rho_0] \notag \\
  &=e^{iH_0^S(n-1)\Delta}\,\mathcal{D}_1\left[e^{-iH_0^S(n-1)\Delta}
    \rho_0\,e^{iH_0^S(n-1)\Delta}
  \right]\,e^{-iH_0^S(n-1)\Delta}, \label{eq:Drel}
\end{align}
where we have defined $\mathcal{D}_0=0$.  Let us denote the state at
$t_0+n\Delta$ as $\tilde\rho_n=\tilde\rho(t_0+n\Delta)$.  

The
perturbative solution~\eqref{eq:evo} is represented as
\begin{equation}
 \tilde\rho_n=C+(\lambda^2\Delta)\,\mathcal{D}_n[C],
\quad C=\tilde\rho_0=\tilde\rho(t_0).
 \label{eq:solC}
\end{equation}
As the state in the interaction picture does not evolve in
$O(\lambda^0)$, $C$ is a constant of motion in this order. Hence, we
regard the $O(\lambda^2)$ term in \eqref{eq:solC} as the secular term due
to the interaction and apply the renormalization group method to
remove the secular behaviour of the solution.  We
introduce an arbitrary renormalization point $k\,(0\le k\le n)$. The
renormalized constant $C_k$ and the counter term $\del C_{k}$ are
introduced as $C=C_k+\del C_{k},\,C_0=C,\,\del C_{0}=0$.
The counter term is chosen so as to absorb the secular term of the solution:
\begin{equation}
  \del C_{k}=C_0-C_k\equiv -(\lambda^2\Delta) 
    \mathcal{D}_k[C_0].       \label{eq:rg0}
\end{equation}
By using the renormalized constant, the naive perturbative solution
\eqref{eq:solC} can be written as
\begin{equation}
  \tilde\rho_n=C_k+(\lambda^2\Delta)(\mathcal{D}_n-\mathcal{D}_k)[C_0],
  \quad (0\le k\le n). \label{eq:rsol}
\end{equation}
Equation~\eqref{eq:rg0} defines a map $\mathcal{R}_k: C_0 \longmapsto
C_k\,(0\le k)$. 
As the renormalization point is arbitrary, the following two equations
hold for two different renormalization points $k_1,k_2 \,(0\le
k_1,k_2\le n)$:
$$
  C_{k_1}-C_0=(\lambda^2\Delta)(\mathcal{D}_{k_1}-\mathcal{D}_0)[C_{0}],\qquad
  C_{k_2}-C_0=(\lambda^2\Delta)(\mathcal{D}_{k_2}-\mathcal{D}_0)[C_{0}].
$$
By subtracting these equations side by side, we obtain the following
relation up to $O(\lambda^2)$:
\begin{equation}
  C_{k_2}-C_{k_1}
    =(\lambda^2\Delta)\left(\mathcal{D}_{k_2}-\mathcal{D}_{k_1}\right)[C_{k_1}],
    \quad (0\le k_1\le k_2).
    \label{eq:rg}
\end{equation}
Thus, this equation defines a map
$\mathcal{R}_{k_2-k_1}: C_{k_1} \longmapsto C_{k_2}\,(0\le k_{2}-k_1)$.
As the map defined by \eqref{eq:rg} satisfies the composition law
$\mathcal{R}_{k_2-k_1}\circ\mathcal{R}_{k_1}=\mathcal{R}_{k_2}
\,(k_1\le k_2)$,
it generates a discretized version of the dynamical
semigroup. Therefore, the evolution of the system can be Markovian in
the discretized (coarse grained) time steps.  In Eq.~\eqref{eq:rsol},
as the renormalization point $k$ is arbitrary, we can choose $k=n$:
\begin{equation}
    \label{eq:rsol2}
    \tilde\rho_n=C_n.
\end{equation}
This is the renormalized solution and the renormalized constant $C_n$
satisfies the discretized version of the renormalization group equation
\eqref{eq:rg}. Combining \eqref{eq:rg} with \eqref{eq:rsol2} and using
the relation \eqref{eq:Drel}, we obtain the
following difference equation for $\tilde\rho_n$:
\begin{align}
 \tilde\rho_n-\tilde\rho_{n-1}&=(\lambda^2\Delta)(\mathcal{D}_n-\mathcal{D}_{n-1}
 )[\tilde\rho_{n-1}] \notag \\
  &=(\lambda^2\Delta)\, e^{iH_0^S(n-1)\Delta}\mathcal{D}_1\left[
    e^{-iH_0^S(n-1)\Delta}\tilde\rho_{n-1}\,e^{iH_0^S(n-1)\Delta}\right]
  e^{-iH_0^S(n-1)\Delta}. \label{eq:diff}
\end{align}
Now, we introduce a new time variable $\tau=\lambda^2 t$ and take the
weak coupling limit $\lambda\rightarrow 0$. For $\tilde\rho(t)$, we
define the time coarse grained state as
\begin{equation}
 \tilde\rho^{\text{CG}}(\tau_n)=\left.\tilde\rho(\tau_n/\lambda^2)
 \right|_{\lambda\rightarrow
   0}=\left.\tilde\rho_n\right|_{\lambda\rightarrow 0}.
\end{equation}
To determine the evolution for the finite interval of time
$\tau_n-\tau_0=\lambda^2n\Delta$, we must take $n\rightarrow\infty$.
In this limit,
\begin{equation}
  \lim_{\lambda\rightarrow 0}\frac{\tilde\rho_n-\tilde\rho_{n-1}}{\lambda^2\Delta}
  =\lim_{\lambda\rightarrow 0}\frac{\tilde\rho^\text{CG}(\tau_{n})
    -\tilde\rho^\text{CG}(\tau_{n}-\lambda^2\Delta)}
  {\lambda^2\Delta}=
  \frac{d}{d\tau}\tilde\rho^\text{CG}(\tau),
\end{equation}
and
$$
\lim_{\lambda\rightarrow
  0}\tilde\rho_{n-1}=\lim_{\lambda\rightarrow 0}
\tilde\rho^{\text{CG}}(\tau_n-\lambda^2\Delta)
=\tilde\rho^{\text{CG}}(\tau_n),\quad
(n-1)\Delta=\lim_{\lambda\rightarrow
  0}\frac{1}{\lambda^2}(\tau_n-\lambda^2\Delta-\tau_0)
 =(\tau_n-\tau_0)/\lambda^2.
$$
Therefore, we obtain the differential equation for the state
$\tilde\rho^{\text{CG}}(\tau)$:
\begin{equation}
    \frac{d}{d\tau}\tilde\rho^{\text{CG}}(\tau)
    = e^{iH_0^S(\tau-\tau_0)/\lambda^2}\mathcal{D}_1\left[
    e^{-iH_0^S(\tau-\tau_0)/\lambda^2}\tilde\rho^{\text{CG}}(\tau)\,
    e^{iH_0^S(\tau-\tau_0)/\lambda^2}\right]e^{-iH_0^S(\tau-\tau_0)/\lambda^2}.
  \label{eq:master0}
\end{equation}
The difference equation \eqref{eq:diff} is converted to the
differential equation in the rescaled time variable $\tau$ with the
small coupling limit. Hence, \eqref{eq:master0} represents the evolution of
the time coarse grained state.

Returning back to the Schr\"{o}dinger representation using
$\tilde\rho^{\text{CG}}(t)=e^{iH_0^S(t-t_0)}\rho^{\text{CG}}(t)e^{-iH_0^S(t-t_0)}$,
we obtain the following linear Markovian master equation:
\begin{equation}
    \frac{d}{dt}\rho^{\text{CG}}(t)=-i[H_0^S,\rho^{\text{CG}}(t)]
    +\mathcal{D}_1[\rho^{\text{CG}}(t)],
    \label{eq:master}
\end{equation}
where the bath-dependent contribution
$\mathcal{D}_1[\rho^{\text{CG}}(t)]$ contains both a Hamiltonian and a
dissipative term
\begin{equation}
    \mathcal{D}_1[\rho^{\text{CG}}(t)]
    =-i[H_{12}^{\Delta},\rho^{\text{CG}}(t)]+\mathcal{L}^{\Delta}
    [\rho^{\text{CG}}(t)], 
\end{equation}
with
\begin{align}
  H_{12}^\Delta&= -\frac{i}{2}\sum_{B_1,B_2}H_{B_1\!B_2}^\Delta
  \,\sigma_{B_1}\sigma_{B_2}, \\
 \mathcal{L}^\Delta[\rho]&=\sum_{B_1,B_2}C_{B_1\!B_2}^\Delta\left(
    \sigma_{B_2}\rho\,\sigma_{B_1}-\frac{1}{2}\left\{
      \sigma_{B_1}\sigma_{B_2},\rho\right\}\right).
\end{align}
Thus, the master equation is
\begin{equation}
  \label{eq:master2}
  \frac{d}{dt}\rho^{\text{CG}}(t)=-i[H_0^S+H_{12}^\Delta,\rho^{\text{CG}}(t)]
  +\mathcal{L}^\Delta[\rho^{\text{CG}}(t)].
\end{equation}
The time-independent coefficients
$H_{B_1\!B_2}^\Delta, C_{B_1\!B_2}^\Delta$ are given by \eqref{eq:HBB}
and \eqref{eq:CBB} replacing $\Delta t\rightarrow\Delta$.  These
coefficients do not contain the initial time and the initial state and
depend only on the coarse graining time $\Delta$.  Therefore, if the
coefficients $C_{B_1\!B_2}^\Delta$ form a positive matrix, the master
equation \eqref{eq:master} has the GKLS form and the state evolves
preserving the trace and complete positivity. 
To show the positivity of the matrix $C_{B_1\!B_2}^\Delta$ defined by
(26), we must check the inequality
$\sum_{B_1\!B_2}\chi_{B_1}^*\,C_{B_1\!B_2}^\Delta\,\chi_{B_2}\ge 0$ for an
arbitrary vector $\chi_B$. The form of the matrix is
\begin{align}
  C_{B_1\!B_2}^\Delta&=\frac{\lambda^2}{\Delta}\left\langle
                     \sum_{A_1}\int_0^\Delta
                     ds_1\Phi_{A_1}(s_1)\,u_{A_1\!B_1}(s_1)
                     \sum_{A_2}\int_0^\Delta
                     ds_2\Phi_{A_2}(s_2)\,u_{A_2\!B_2}(s_2)\right\rangle \notag\\
  &\equiv\frac{\lambda^2}{\Delta}\left\langle L_{B_1}L_{B_2}\right\rangle,
\end{align}
where we introduced the operator
\begin{equation}
 L_B\equiv\sum_A\int_0^\Delta\Phi_A(s) u_{AB}(s).
\end{equation}
Without loss of generality, we can assume that the operators $\Phi_A$ and
$\sigma_A$ are Hermitian. Then the function $u_{AB}(t-t_0)$ representing
the evolution of the system variable $\sigma_A$ is real and the
operator $L_A$ is the Hermitian $L_B=L_B^\dagger$. Hence,
\begin{align}
  \sum_{B_1B_2}\chi^*_{B_1}C_{B_1\!B_2}^\Delta\chi_{B_2}&=\frac{\lambda^2}{\Delta}
                                                        \sum_{B_1,B_2}
                     \left\langle\chi_{B_1}^*L_{B_1}L_{B_2}\chi_{B_2}
                                                        \right\rangle\notag\\
  &=\frac{\lambda^2}{\Delta}\left\langle\left(\sum_{B_1}\chi_{B_1}L_{B_1}\right)
    ^\dagger\left(\sum_{B_2}\chi_{B_2}L_{B_2}\right)\right\rangle \notag\\
  &\ge 0,
\end{align}
because this quantity is an expectation value of the positive
operator. Therefore,  the coefficients $C_{B_1\!B_2}^\Delta$  form a
positive matrix.
The master equation \eqref{eq:master2} is
the same one derived in Refs.~5 and 12-14.

\section{Summary}

In this paper, by the renormalization group method, we have
rederived the Markovian master equation in the GKLS form under the
following assumptions:
  \begin{enumerate}
    \item The interaction between the system and the bath is weak (weak
    coupling) and the back action of the system on the bath is
    negligible. 
    \item The correlation time of the bath field is sufficiently
    shorter than the relaxiation time of the system and the existence
    of the factorized state at the specific initial time $t_0$ can be
    assumed.
    \item The coarse graining time $\Delta$ is longer than the time
    scale of the bath field.
  \end{enumerate}
  The naive perturbative solution Eq.~\eqref{eq:naive2} does not have
  a form of the master equation. To transform it to the master
  equation, we introduced the coase graining time scale $\Delta$,
  which is larger than the bath time scale $t_B$. This reduces the
  double integral with respect to time in the solution to the single
  integral, and this reduction is equivalent to introducing the
  Markovian approximation carried out in conventional derivations of
  the master equation.  However, at this stage, the equation does not
  have the form of the master equation.  Then, we applied the
  renormalization group method and eliminated a specific initial
  time. Conventional derivations of the master equation usually do the
  same thing by requiring the time translational invariance of the
  considering system and the structure of the equation does not depend
  on the specific initial time. We have carried out the equivalent
  procedure by assuming that the structure of the evolution does not
  depend on the specific renormalization point and derived the
  renormalization group equation, which provides the desired master
  equation.

  The obtained renormalization group equation generates a dynamical
  semigroup for the renormalized constant, and the renormalized
  density operator for the slow time variable $\tau$ satisfies the
  time local Markovian master equation in Lindblad form. As discussed
  in Refs.~5 and 12-14, the master equation \eqref{eq:master2} reduces
  to the one with the rotational wave approximation and the secular
  approximation in the limit $\Delta\rightarrow\infty$. However, for
  the purpose of detection of the entanglement of the quantum field
  using particle detectors, the master equation with the time coarse
  graining is more suitable because it can detect the vacuum quantum
  fluctuation, which is the cause of the entanglement of the quantum
  field.

\begin{acknowledgments}
 This work was supported in part by  JSPS Grant-in-Aid for
 Scientific Research (C) (23540297).
\end{acknowledgments}


\begin{thebibliography}{10}
\bibitem{foot}
We assume homogeneity of the bath
  correlation function; it only depends on the difference of times.
\end{thebibliography}

\end{document}